\documentclass[english,twocolumn,aps,pra,showpacs]{revtex4}
\usepackage[T1]{fontenc}
\usepackage[latin1]{inputenc}
\usepackage{graphicx}
\usepackage{amssymb}

\makeatletter

\input{epsf}
\usepackage{babel}

\makeatother

\begin{document}

\title{Josephson effect in an atomic Fulde-Ferrell-Larkin-Ovchinnikov superfluid}

\author{Hui Hu$^{1}$ and Xia-Ji Liu$^{1}$}

\affiliation{$^{1}$\ ARC Centre of Excellence for Quantum-Atom Optics, Centre
for Atom Optics and Ultrafast Spectroscopy, \\
Swinburne University of Technology, Melbourne 3122, Australia}

\date{\today{}}

\begin{abstract}
We study theoretically two spatially separate quasi-one-dimensional atomic 
Fermi gases in a double-well trap. By tuning independently their spin polarizations, 
a Fulde-Ferrell-Larkin-Ovchinnikov (FFLO) superfluid or a Bardeen-Cooper-Schrieffer 
(BCS) superfluid may be formed in each well. We seek the possibility of creating 
a spatially tunable atomic Josephson junction between two superfluids, which is 
supposed to be realizable via building a weak link at given positions of the 
double-well barrier. We show that within mean-field theory the maximum Josephson 
current is proportional to the order parameter in two wells. Thus, the spatial 
inhomogeneity of the FFLO order parameter in one well may be directly revealed 
through the current measurement with the position-tunable link. We anticipate that 
this type of Josephson measurements can provide a useful evidence for the existence 
of exotic FFLO superfluids. Possible experimental realizations of the Josephson 
measurements in atomic Fermi gases are discussed.
\end{abstract}

\pacs{03.75.Ss, 05.30.Fk, 71.10.Pm, 74.20.Fg}

\maketitle

\section{Introduction}

Strongly attractive Fermi gases with imbalanced spin components
are ubiquitous systems in diverse fields of physics \cite{rmp}. They are
building-blocks of atomic nuclei, the matter in neutron stars and even the
quark-gluon plasma that comprised the early Universe. Imbalanced Fermi gases
also appear in solid-state superconductors subjected to either an internal exchange
field or external magnetic field. A recent example attracted intense attentions 
is the trapped atomic gases of neutral fermions with unequal or polarized spin 
populations \cite{mit,rice}. Owing to the flexibility in the control of the 
constituents and interaction strengths, atomic Fermi gases provide the most 
promising place for observing many exotic forms of matter. 

The ground state of polarized Fermi gases remains elusive \cite{hl}. The mismatched Fermi 
surfaces in polarized environment cannot guarantee the standard Bardeen-Cooper-Schrieffer 
(BCS) mechanism, which requires a pairing of two fermions on the same Fermi surface 
with opposite spins. Various exotic forms of pairing have been suggested \cite{fflo,dfs,wvliu,sarma,bedaque}, 
such as Fulde-Ferrell-Larkin-Ovchinnikov (FFLO) state with spatially varying 
order parameters \cite{fflo}, deformed Fermi surface \cite{dfs}, interior gap \cite{wvliu} 
or Sarma superfluidity \cite{sarma}, and phase separation \cite{bedaque}.

Among these, the FFLO state is of particular interest, since the Cooper pairs
may condense into a state with a finite center-of-mass momentum. The search for
the FFLO state has lasted for more than four decades in many branches
of physics. In condensed matter community, experimental evidences of 
its existence have been reported in the heavy fermion superconductor CeCoIn$_5$. 
The observations include the specific heat, magnetization, and penetration depth 
measurements \cite{cecoin5}. Recent theoretical studies suggest that such phase 
is more favorable in the low-dimensional systems \cite{yang,orso,hld,parish,feiguin,xiajipra1d,guan1,xiajimulti1d,zhao,casula,xiajipra1d2008,luscher,rizzi,gao,batroun,tezuka,guan2,kakashvili,wang,chen,guan3,edge}.
Particularly it dominates in the quasi-one-dimensional (1D) polarized gases~\cite{orso,hld}. 
Following these suggestions, most recently, strong evidence 
for the FFLO superfluid has been found in ultracold atoms at Rice University \cite{riceexpt}, 
by trapping a two-component mixture of ultracold $^6$Li atoms in an array of one dimensional tubes. 
At temperatures $T\sim 0.1T_F$, where $T_F$ is the Fermi temperature, the measured 
density profiles exhibit a partially polarized core surrounded by wings 
composed of either a completely paired BCS superfluid or a fully polarized normal gas, 
in excellent agreement with the theoretical predictions given by Orso \cite{orso} and 
by the present authors \cite{hld}. 

However, the Rice experiment was done with an intermediate interaction strength, 
where the 1D binding energy is much larger than the Fermi energy. Therefore, more 
accurate measurements are required at weaker interactions (and hence lower temperatures), 
together with a new definitive identification scheme for the FFLO order parameter. 
These should be based on phase-sensitive measurements that can directly reveal 
the spatial variations of the phase of the order parameter. One possibility is the 
Josephson effect \cite{je}.

In this work we propose an \emph{atomic Josephson effect} to 
detect the existence of the exotic FFLO superfluids. We consider two spatially separate quasi-1D atomic Fermi gases with 
tunable spin polarizations in a tight double-well potential, where the lateral 
motion of fermions is frozen, while axial motion is weakly confined. A weak link 
at a specific position $x_0$ may be created by superimposing a narrow dipole 
dimple potential to allow tunneling. Fig. 1 presents a schematic view of the 
configuration, together with the potential and particle density profiles. Its 
possible realization will be addressed later.

\begin{figure}

\begin{center}
\includegraphics[clip,width=0.48\textwidth,angle=0]{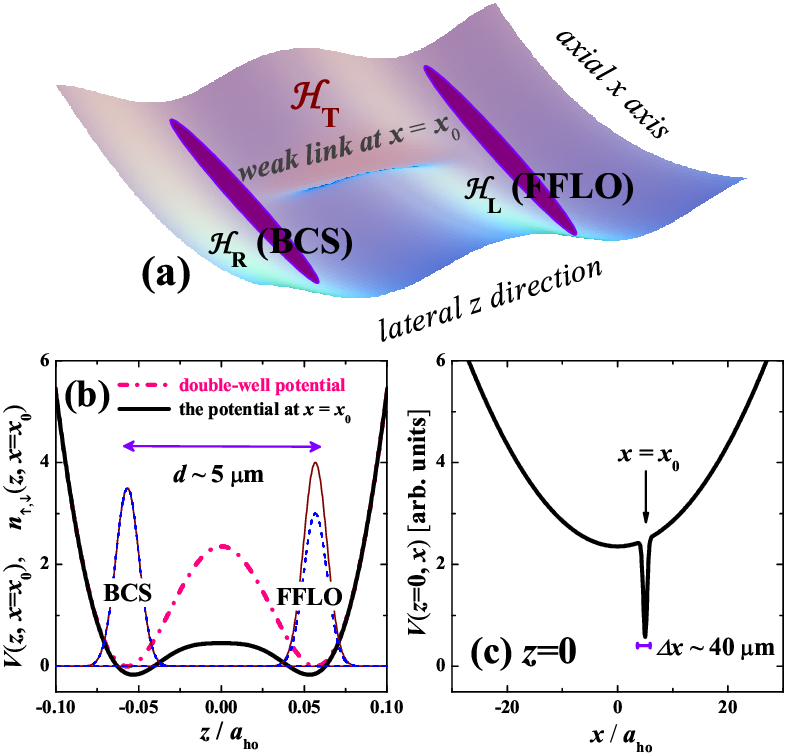}
\par\end{center}

\caption{(Color online) Schematic view of a proposed atomic Josephson junction in quasi-1D 
atomic Fermi gases with a number of total atoms $N \sim 10^3$. (a) The double-well potential 
landscape. Two needle-like (BCS and FFLO) superfluids are located in the left and right 
wells, respectively. Reducing the double-well barrier at a specific position $x=x_0$ by 
superimposing a narrow dipole dimple potential, the Cooper pairs in superfluids can 
tunnel back and forth between wells, leading to a Josephson current. (b) Lateral 
distributions of the potential (red thick dot-dashed line) and spin-up (red thin solid line) 
and spin-down (blue dashed line) particle density profiles away from the weak link at $x_0$. The 
inter-well distance in experiments would be about $5\mu m$. The potential with 
superposition of a dimple potential is shown by the black thick solid line. Then, the particle 
densities and the order parameters can have appreciable overlaps within the weak link. The 
length scale $a_{\text{ho}}=\sqrt{\hbar /m\omega }$ is around $40\mu m$ for $^{6}$Li atoms, 
where $\omega$ is the trapping frequency in the axial direction. (c) Axial profile 
of the "harmonic" + "dimple" potential at $z=0$.}

\label{fig1} 
\end{figure}

\begin{figure}
 
\begin{centering}
\includegraphics[clip,width=0.48\textwidth]{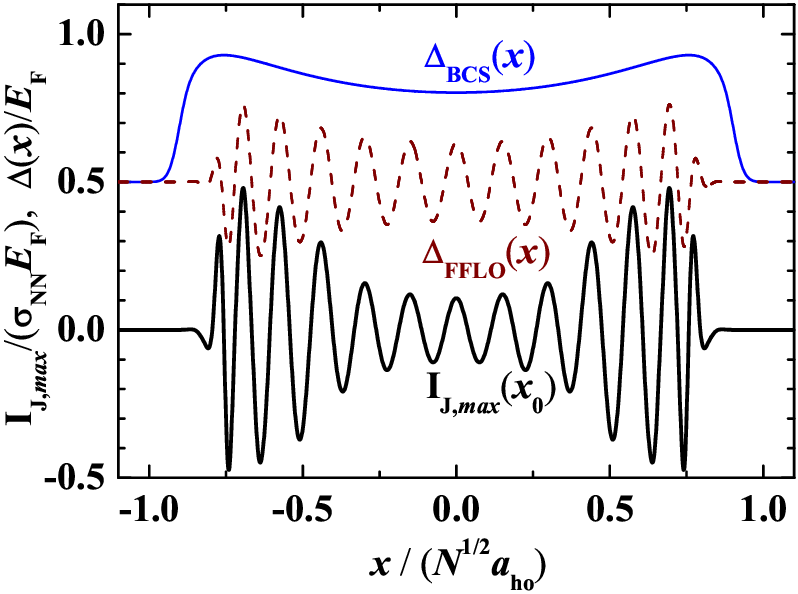}
\par\end{centering}

\caption{(Color online) Underlying physics of the Josephson effect, as a probe of the exotic 
FFLO superfluids. With a BCS and a FFLO superfluid placed on the left and right sides 
of the junction, respectively, the calculated maximum Josephson current is roughly 
proportional to the FFLO order parameter. We have chosen the number of fermions on each well $N_L=N_R=128$,
compared to the realistic number of $N \sim 10^3$. The interaction strengths are $\gamma _L=$ $\gamma _R=1.6$. 
The order parameter is normalized by the Fermi energy $E_F=N_L\hbar \omega /2$, and is 
shifted upwards by an amount of $0.50E_F$ for clarity. The spin polarization in 
the FFLO superfluid is $0.25$. $\sigma _{NN}$ is the conductance of a corresponding 
normal junction. See, for example, the text in Sec. IIIA.}

\label{fig2}
\end{figure}

The underlying physics of our proposal is easily understood using Ginsburg-Landau (GL) 
theory~\cite{BdG}. Assuming that the order parameters or condensate wave functions in 
the left and right wells are described by $\Psi _{\text{BCS}}(x)$ and $\Psi _{\text{FFLO}}(x)$, 
respectively, the Josephson current by GL theory is 
\begin{equation}
I_J=%
\mathop{\rm Im}%
\left[ J\int_{x_0-\Delta x/2}^{x_0+\Delta x/2}dx\Psi _{\text{BCS}}^{*}(x)\Psi _{\text{FFLO}}(x)\right] ,  \label{GL}
\end{equation}
where $J$ is the characteristic tunneling amplitude and is determined by the 
small overlap of the condensate wave functions along the lateral direction. 
$\Delta x$ is the width of the narrow tunneling link at position $x_0$. 
The BCS order parameter in Eq. (\ref{GL}) is essentially spatially independent, 
while the FFLO one is oscillatory in real space with period $2\pi\hbar /q_{\text{FFLO}}$, 
where $q_{\text{FFLO}}$ is the center-of-mass momentum. Thus, provided that 
$\Delta x\ll 2\pi \hbar /q_{\text{FFLO}}$, the measurement of the maximum 
Josephson current results directly $\Psi _{\text{FFLO}}(x=x_0)$. By displacing 
axially the harmonic traps and consequently changing the position of weak 
link $x_0$, a series of measurements therefore reveal the whole spatial 
inhomogeneity of the FFLO order parameter. As shown in Fig. 2, the simple 
GL picture is verified by much complicated microscopic calculations, which 
will be outlined in detail below.

Further manipulation of temperature, number of total atoms, or interaction strengths 
via a Feshbach resonance may lead to an \emph{atomic superfluid-normal junction}. In this
case, the difference in chemical potentials between wells, resulting from the different total 
number of atoms, plays the role of the voltage. Therefore, analogous to the differential conductance
measurements in superconductors~\cite{BdG}, the derivative of
single-particle tunneling currents with respect to the number difference
provides a direct measurement of the density of states (DOS) of superfluids. 
We show that the characteristic two-energy-gap structure in the DOS of the FFLO 
state can be clearly determined, giving an independent means of 
identifying its existence.

Our results are obtained by solving mean-field Bogoliubov-de Gennes
(BdG) equations for each well, while treating the tunneling through the weak
link within linear response theory. Our calculations are performed
specifically for atomic Fermi gases. However, as the FFLO physics is a
fundamental issue that is of importance to many research fields,
they can have potential implications beyond ultracold atoms.

\section{Theoretical model}

We assume that the dipole dimple potential is a small perturbation 
and the resulting weak link at $x_0$ does not disturb considerably the 
distributions of the order parameter and particle density profile in each well. 
The atomic Josephson junction in Fig. 1a then is well described by a 
tunneling Hamiltonian. By integrating out the lateral degree of freedoms 
in a tight-binding approximation, it takes three terms:
\begin{eqnarray}
{\cal H} &=&{\cal H}_L+{\cal H}_T+{\cal H}_R,  \label{hami} \\
{\cal H}_T &=&V_0\sum\nolimits_\sigma \left[ \psi _{L\sigma }^{+}\left(
x_0\right) \psi _{R\sigma }\left( x_0\right) +H.c.\right] ,
\end{eqnarray}
where $\sigma =\uparrow ,\downarrow $ is the spin index. The terms ${\cal H}%
_L$ and ${\cal H}_R$ are respectively the Hamiltonians for fermions on the
left and right sides of the junction, and can be expressed in
terms of operators $\psi _{L,R;\sigma }\left( x\right) $.
They contain all the many-body interactions. Assuming that the width $\Delta x$ 
is the smallest length scale, in ${\cal H}_T$ we approximate all operators 
$\psi \left( x\right) \approx \psi \left(x_0\right) $ and introduce a transfer 
parameter $V_0=J\Delta x$. This is valid as far as $\Delta x\ll 2\pi \hbar /q_{\text{FFLO}}$.
We shall take a small constant transfer parameter, which
corresponds to the small overlap of two order parameters. The overlap could 
depend weakly on the position of the weak link. However, the assumption of a 
fixed transfer parameter is sufficient to capture the qualitative feature of 
the Josephson effect. 

In each well the ground state of an attractive gas of $N=N_{\uparrow
}+N_{\downarrow }$ fermions with polarization $P=(N_{\uparrow
}-N_{\downarrow })/N$ is conveniently determined by using the BdG formalism~\cite{BdG} 
that describes the quasiparticle wave functions $u_\eta (x)$ and $v_\eta (x)$ with 
a contact interaction $g$ (the well index is suppressed for clarity), 
\begin{equation}
\left[ 
\begin{array}{cc}
{\cal H}_{\uparrow }^0-\mu _{\uparrow } & \Delta (x) \\ 
\Delta ^{*}(x) & -{\cal H}_{\downarrow }^0+\mu _{\downarrow }
\end{array}
\right] \left[ 
\begin{array}{c}
u_\eta \left( x\right) \\ 
v_\eta \left( x\right)
\end{array}
\right] =E_\eta \left[ 
\begin{array}{c}
u_\eta \left( x\right) \\ 
v_\eta \left( x\right)
\end{array}
\right] ,  \label{BdG}
\end{equation}
where ${\cal H}_{\uparrow ,\downarrow }^0=-\hbar ^2{\bf \nabla }%
^2/2m+m\omega ^2x^2/2+g_{1d}n_{\downarrow ,\uparrow }(x)$ is the single particle
Hamiltonian under axial harmonic trap and Hartree potential. The chemical
potentials are shifted as $\mu _{\uparrow ,\downarrow }=\mu \pm \delta \mu $ 
to account for the unequal population $N_{\uparrow ,\downarrow }$. The order parameter
$\Delta (x)$ and $\mu _{\uparrow ,\downarrow }$ are calculated by self-consistency 
equations for the gap, $\Delta (x)=g_{1d}\sum_\eta u_\eta (x)v_\eta ^{*}(x)f(E_\eta )$, 
and for the densities: 
$n_{\uparrow }(x)=\sum_\eta \left| u_\eta \left( x\right)\right| ^2f(E_\eta )$ and 
$n_{\downarrow }(x)=\sum_\eta \left| v_\eta(x)\right| ^2f(-E_\eta )$, 
with $f(x)=1/(\exp [x/k_BT]+1)$ being the Fermi function. These must be 
constrained so that $\int dxn_{_{\uparrow ,\downarrow }}\left( x\right) =N_{\uparrow ,\downarrow }$.
We note that the unequal chemical potentials break the time-reversal
symmetry. Thus, the sum over energy levels is done for all the eigenstates
with both positive and negative energies $E_\eta $. Generically,
the interaction strength is parameterized by a dimensionless coupling constant 
$\gamma =-mg_{1d}/(\hbar ^2n_0)$, where $n_0$ is the center density of
an ideal gas. In the weak or intermediate coupling regimes (\textit{i.e.}, 
$\gamma \lesssim 10^0$), the mean-field BdG theory appears to be very
accurate. More rigorous treatment should include the pair fluctuations beyond mean-field \cite{ournsr1,ournsr2,ournsr3,ournsr4,ournsr5}.
Fig. 3 shows the mean-field results of density profiles of a
gas at $\gamma =1.6$ and $P=0.05$, $0.25$, compared with that obtained from exact 
Gaudin solutions and local density approximation~\cite{hld,xiajipra1d}. 
The agreement is reasonable. Both theories predict two-shell structures with a 
partially polarized superfluid at the trap center and either a fully paired (small $P$) 
superfluid or a fully polarized (large $P$) normal state at the edge. The mean-field 
order parameters at these polarizations are given in Fig. 4b (for $P=0.05$) and Fig. 2 (for $P=0.25$), 
respectively. Their spatial variation identifies clearly that the partially 
polarized phase at center is indeed a FFLO superfluid.

\begin{figure}
 
\begin{centering}
\includegraphics[clip,width=0.48\textwidth]{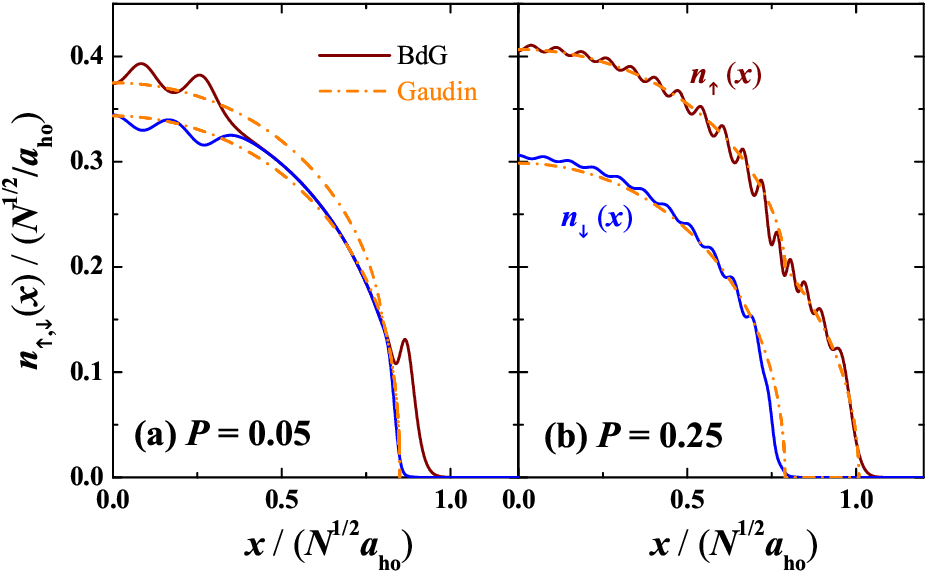}
\par\end{centering}

\caption{(Color online) Zero temperature mean field density profiles (solid lines) 
of a polarized gas at $N=128$ and $\gamma =1.6$, compared with the results from 
Gaudin solutions (dot-dashed lines)~\cite{hld,xiajipra1d}. The oscillation in the 
density profiles, the so-called Friedel oscillation, is due to finite size effect. It becomes
negligibly small for large enough number of atoms. In contrast, the oscillation
in the order parameter (scaled by the Fermi energy), the unambiguous signature 
of the FFLO phase, is robust with respect to the change of the number of total atoms.
We refer to Sec. IVA in Ref. \cite{casula}  and Sec. IIIA in Ref. \cite{xiajipra1d2008} 
for a more detailed discussion of the Friedel oscillation.}

\label{fig3} 
\end{figure}

\section{Josephson effect and single-particle tunneling}

The main observable of interest, the rate of transferred atoms from, \textit{e.g.}, 
right well to left well, is defined by $I(t)=<d\hat{N}_L(t)/dt>$. In
analogy to superconductors where the flow of electrons out of the
superconductor establishes an electrical current, we call $I$ the current.
By rewriting the transfer Hamiltonian ${\cal H}_T=\sum\nolimits_\sigma
(A_\sigma +A_\sigma ^{+})$ where $A_\sigma =\psi _{L\sigma }^{+}\left(
x_0\right) \psi _{R\sigma }\left( x_0\right) $, the equation of motion leads
to $d\hat{N}_L(t)/dt=i[{\cal H}_T(t),\hat{N}_L(t)]=i\sum\nolimits_\sigma
[A_\sigma (t)-A_\sigma ^{+}(t)]$. Bearing in mind that the link at $x_0$ is
weak so that the transfer of atoms can be treated as a perturbation, we use
the linear response theory~\cite{mahan}, in which the current $%
I(t)=-i\int_{-\infty }^tdt^{\prime }<[d\hat{N}_L(t)/dt,{\cal H}_T(t^{\prime
})]>_0=\sum\nolimits_{\sigma \sigma ^{\prime }}\int_{-\infty }^tdt^{\prime
}<[A_\sigma (t)-A_\sigma ^{+}(t),A_{\sigma ^{\prime }}(t^{\prime
})+A_{\sigma ^{\prime }}^{+}(t^{\prime })]>_0$. Here the subscript 0 in the average
refers to the unperturbed systems ${\cal H}_L$ and ${\cal H}_R$. Two contributions 
can be easily identified: the normal single-particle current and the
Josephson current of Cooper pairs. The explicit expression of the latter is
given by~\cite{mahan} $I_J(t)=\exp [i(\varphi _R-\varphi _L)+i2(\mu _R-\mu
_L)t/\hbar ]\sum\nolimits_\sigma \int_{-\infty }^tdt_1\exp [-i(\mu
_{R,-\sigma }-\mu _{L,-\sigma })(t-t_1)/\hbar ]<[\tilde{A}_\sigma (t),\tilde{%
A}_{-\sigma }(t_1)]>_0+c.c.$, where we introduce an interaction representation
with respect to ${\cal H}_L$ and ${\cal H}_R$, and represent it by a tilde 
in operators. The global phases of order parameters, $\varphi _L$ 
and $\varphi _R$, are made explicit in $I_J(t)$. Their difference, 
together with the factor $2(\mu _R-\mu_L)t/\hbar $, drives the direct- and/or 
alternating-current Josephson currents even at the zero chemical potential 
difference between wells.

\subsection{Josephson tunneling}

We first concentrate on the Josephson tunneling with the same number of
particles in each well, for which the single-particle tunneling is blocked. With the help of the 
Wick theorem, in the statistical average of $I_J(t)$ we split the four fermionic 
field operators. The integration of the average over time $t_1$ can then be expressed 
in terms of the retarded correlation functions~\cite{mahan},
\begin{eqnarray}
\chi _{_{\downarrow }{}_{\uparrow }}\left( \Omega \right)
&=&\sum\nolimits_{ij}\left( u_iv_i^{*}\right) _L\left( u_j^{*}v_j\right) _R%
\frac{\left[ f\left( E_i\right) -f(E_j)\right] }{+\Omega +E_i-E_j},
\label{cf1} \\
\chi _{_{\uparrow }{}_{\downarrow }}\left( \Omega \right)
&=&\sum\nolimits_{ij}\left( u_iv_i^{*}\right) _L\left( u_j^{*}v_j\right) _R%
\frac{\left[ f\left( E_i\right) -f(E_j)\right] }{-\Omega +E_i-E_j},
\label{cf2}
\end{eqnarray}
where the indices $i$ and $j$ refer to, respectively, the energy levels
in the left and right wells, and the subscripts $L$ and $R$ are the indices
for wells. We have abbreviated $u=u(x_0)$ and $v=v(x_0)$. Consequently, the
Josephson current can be calculated as~\cite{mahan}, 
\begin{equation}
I_J\left( t\right) =I_J^{\max }\sin \left[ \left( \varphi _R-\varphi
_L\right) +2\left( \mu _R-\mu _L\right) t/\hbar \right] ,  \label{jc}
\end{equation}
where the maximum current 
\begin{equation}
I_J^{\max }=\frac{2V_0^2}\hbar \left[ \chi _{_{\downarrow }{}_{\uparrow
}}\left( \mu _{R\downarrow }-\mu _{L\downarrow }\right) +\chi _{_{\uparrow
}{}_{\downarrow }}\left( \mu _{R\uparrow }-\mu _{L\uparrow }\right) \right] .
\end{equation}
The Josephson current thus oscillates in phase with a peak value $I_J^{\max
} $. It is clear from the expressions (\ref{cf1}) and (\ref{cf2}) that
correlation functions become roughly a product of two order parameters if we
approximate the gap equation $\Delta (x)\propto \sum_\eta u_\eta (x)v_\eta
^{*}(x)$. This particular structure emphasizes the microscopic origin of the
GL theory of Josephson effect.

\begin{figure}
 
\begin{centering}
\includegraphics[clip,width=0.48\textwidth]{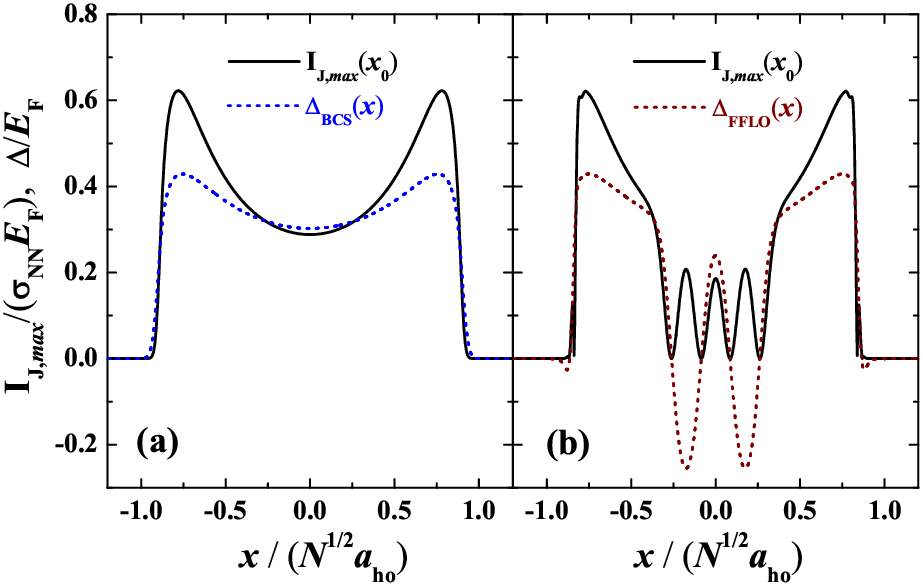}
\par\end{centering}

\caption{(Color online) Maximum Josephson current of a junction with identical
superfluids of either BCS (a) or FFLO (b) type, as a function of the
position of the weak link. The spin polarization in (b) is $0.05$. The order
parameters (dashed lines) are plotted to emphasize their similarity
to the current, \textit{i.e.}, $I_{J,\max }\propto \left| \Delta(x_0)\right| $. 
The other parameters are the same as in Fig. 2.}

\label{fig4} 
\end{figure}

A simple expression for $I_J^{\max }$ can be derived when we assume
identical uniform BCS superfluids in both wells, $I_J^{\max }=(\pi /2)\sigma
_{NN}\Delta $, where $\sigma _{NN}=(V_0^2/\pi )[2m/(\hbar ^3E_F)]$ may be
viewed as the conductance of a normal junction, with $E_F$ being the Fermi
energy. Therefore the value of gap can be determined by measuring $I_J^{\max
}$. In the presence of traps or inhomogeneous superfluids, one has to resort
to the numerical calculations. Fig. 4 presents the maximum Josephson currents 
for identical BCS or FFLO superfluids in both wells. Roughly we find 
$I_{J,\max }\propto \left| \Delta(x_0)\right| $. A more promising scheme is 
provided in Fig. 2, where a BCS superfluid is set in the left well as a 
reference system, while the order parameter of a FFLO superfluid in the right 
well is to be determined. As a result of the flat distribution of BCS order 
parameter, we anticipate $I_J^{\max} \propto \Delta _{\text{FFLO}}(x_0)$, which is confirmed 
numerically. Therefore, the spatial inhomogeneity of FFLO phases can be 
precisely detected by Josephson effect at varying positions of the weak link.
This constitutes the main result of the present paper.

\begin{figure}
 
\begin{centering}
\includegraphics[clip,width=0.40\textwidth]{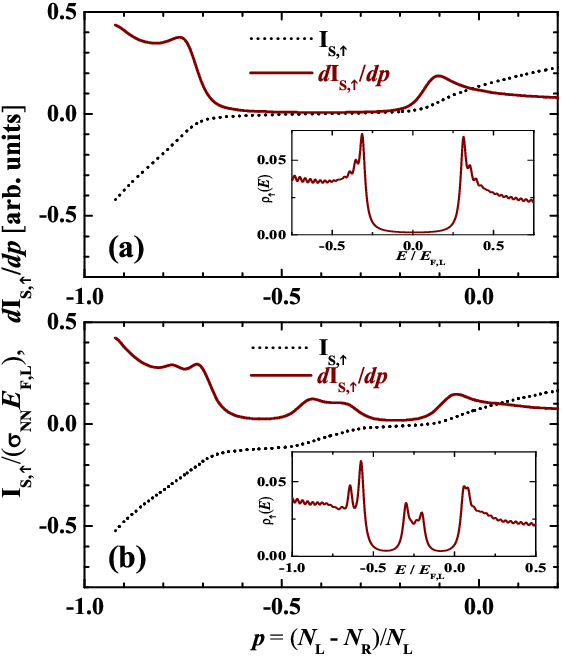}
\par\end{centering}

\caption{(Color online) Spin up single-particle current and its derivative as
a function of the imbalance between wells. An unpolarized normal gas with 
varying number of particles $N_R$ is set in the right well. We put in the
left well a BCS superfluid (a) or a FFLO superfluid at polarization $0.05$ (b), 
at $N_L=128$ and $\gamma _L=1.6$.}

\label{fig5} 
\end{figure}

\subsection{Single-particle tunneling}

We evaluate next the single-particle tunneling current. Again by the use of
correlation functions, one ends up with a familiar expression~\cite{mahan}, $%
I_S=I_{S,\uparrow }(\mu _{R\uparrow }-\mu _{L\uparrow })+I_{S,\uparrow }(\mu
_{R\downarrow }-\mu _{L\downarrow })$, where 
\[
I_{S,\sigma }=\frac{2\pi V_0^2}\hbar \int\nolimits_{-\infty }^{+\infty
}d\epsilon \rho _{L\sigma }(\epsilon +\Omega )\rho _{R\sigma }(\epsilon
)\left[ f\left( \epsilon \right) -f\left( \epsilon +\Omega \right) \right] 
\]
and $\rho _{\uparrow }(\epsilon )=$ $\sum_\eta \left| u_\eta \left(
x_0\right) \right| ^2\delta (\epsilon -E_\eta )$ and $\rho _{\downarrow
}(\epsilon )=$ $\sum_\eta $ $\left| v_\eta \left( x_0\right) \right|
^2\delta (\epsilon +E_\eta )$ are the spin up and spin down DOS at the 
position of the weak link. We are interested in the superfluid-normal 
junction, where the DOS of the well in normal state is essentially a constant. 
Thus, the derivative of currents with respect to the chemical potential 
difference or number difference between wells provides a direct measurement of the DOS
in the superfluid well. This is a phenomenon reminiscent of the scanning
tunneling microscopy--here the role of the tip is played by the normal gas
in one well. Fig. 5 predicts such measurements for BCS and FFLO superfluids 
at the trap center $x_0=0$. All the features in the spin up DOS of superfluids
(insets in the figure) are faithfully recovered in the differential
conductance $dI_{S,\uparrow }/dp$, which is calculated by changing the
relative number difference between wells $p=(N_L-N_R)/N_L$. In particular, the
midgap state or two-energy-gap structure in the FFLO DOS, a salient feature 
due to the spatially modulated order parameter, is clearly visible in 
$dI_{S,\uparrow }/dp$. This presents an independent check of the existence 
of FFLO phases.

\section{Possible experimental realizations of the atomic Josephson junction}

We are now in position to discuss the experimental realization of the atomic
Josephson effect. The major experimental challenge is the reach of a 
quasi-1D fermionic superfluid at the lowest experimentally accessible 
temperature, which is about $0.05T_F$. To have reasonable superfluid transition 
temperatures, a feasible way of tuning the inter-atomic interactions is required, such 
as Feshbach resonances. On the other hand, an optical lattice 
may be used to effectively reduce the dimensionality of atomic Fermi gases.
 
Thus, we start from a fermionic $^6$Li gas with a number of total atoms 
$N \sim 10^3$ in an optical dipole trap, which was realized recently by the Hulet 
group at Rice University~\cite{rice}. The optical trap is highly elongated, 
with an aspect ratio of radial to axial trapping frequencies 
$\omega_{\bot}/\omega \sim 50$~\cite{rice}, Then, we consider the superposition 
of a deep periodic potential along the radial (lateral) direction. A double-well 
configuration will be formed. By suitably choosing the depth and periodicity 
of the optical lattice, the aspect ratio may increase to several hundreds or 
up to a thousand. This technique has already been applied successfully to 
create a bosonic Josephson junction for $^{87}$Rb gases with a similar number 
of total atoms, but in a much less anisotropic optical trap~\cite{albiez}. 
Next, a narrow weak link between wells may be built up by adding a tight 
dipole dimple microtrap~\cite{dimple} at a specific position. The position of the link
could be easily to displace. We note that prepared in this way 
the typical inter-well distance will be about several micrometers. For such a 
short distance, it is difficult to manipulate independently the spin polarization in 
each well by a radio-frequency sweep~\cite{mit,rice}. In this respect, the Josephson 
measurement between two FFLO superfluids, as shown in Fig. 4b, seems to be 
more feasible, compared to the proposal outlined in Figs. 1 and 2.  

We turn to estimate some realistic experimental parameters. To observe the 
Josephson oscillations, it is necessary to fulfill two conditions: (i) the number of 
fermions involved in the oscillations, to be measured by phase-contrast imaging \cite{mit,rice}, 
should be large enough to be easily detected, but small
enough to ensure the validity of the linear response theory; and (ii) the
width of the weak link $\Delta x$ should be much smaller than the period of
FFLO order parameter $2\pi \hbar /q_{\text{FFLO}}$. 

Typically, the total number of atoms in one well would be around $N\sim 10^3$, and 
the frequency of axial trap $\omega \sim 2\pi \times 1$ {\rm Hz}, which is much 
smaller than the radial frequency $\omega _{\bot }\sim 2\pi \times 10^3$ {\rm Hz} so
that the quasi-1D condition $N\omega \leq \omega _{\bot }$ nearly holds~\cite{schumm}.
The tunnling barrier is on the order of $\hbar \omega_{\bot}$. Further, we select 
the total spin polarization $P=0.05$. For $^{6}$Li atoms, with these parameters we find 
$2\pi \hbar /q_{\text{FFLO}} \sim 2\pi / (k_{\uparrow}-k_{\downarrow}) \sim 2\pi/(N^{1/2}P)a_{\text{ho}} \sim 160$ $\mu ${\rm m},
where $a_{\text{ho}}=\sqrt{\hbar /m\omega } \sim 40$ $\mu ${\rm m} 
is the harmonic oscillator length along the axial direction. This FFLO period is 
much larger than the width of dimple potential $\Delta x$ that is about several 
ten micrometer, and therefore the condition $\Delta x\ll 2\pi \hbar /q_{\text{FFLO}}$ 
can be well satisfied. Roughly, at the trap center there are about 100 fermions 
in each FFLO period. Choosing $\Delta x \sim 40$ $\mu ${\rm m}, we expect 
about 25 particles on average on one side of the tunneling window in each well.

The maximum Josephson current $I_{J,\max}$ depends critically on the weak link at
position $x_0$. To estimate it, we quote the parameters from previous Josephson 
measurements for bosonic $^{87}$ Rb atoms, $I_{J} \sim 5 \times 10^4 ~\text{sec}^{-1}$ 
in Ref.~\cite{albiez} and $I_{J} \sim 3 \times 10^6 ~\text{sec}^{-1}$ in 
Ref.~\cite{levy}. Due to the use of a narrow dimple potential, the maximum Josephson 
current in our configuration should be much reduced. It is not unreasonable to estimate 
$I_{J,\max } \sim 10^3 ~\text{sec}^{-1}$, whose value is two or three orders of magnitude 
smaller than that of bosonic Josephosn junctions~\cite{albiez,levy}. On the other hand, 
assuming a time scale of oscillation $\sim 0.05$ {\rm sec}~\cite{albiez}, the number of 
transferred fermions is about 50, an order of magnitude smaller than the total number 
of atoms. This number may be within the present experimental detection limits, 
\textit{i.e.}, the Josephson oscillation in a Bose-Einstein condensate with similar 
number of atoms and time scale has already been observed \cite{albiez}.

It is worth noting that the proposed atomic Josephson experiment requires a series of 
measurements (shots) with varying position of the weak link. Thus, the experiment 
relies on a good repeatability from shot to shot. For example, the nodes and anti-nodes 
of the oscillating FFLO order parameter should always occur at the same point in the 
cloud. However, the node of the FFLO wave-functions may be sensitive to some experimental 
imperfections such as fluctuations in number imbalance and in temperature. This 
would wash out the FFLO signal of the measurement.

\begin{figure}
\begin{centering}
\includegraphics[clip,width=0.40\textwidth]{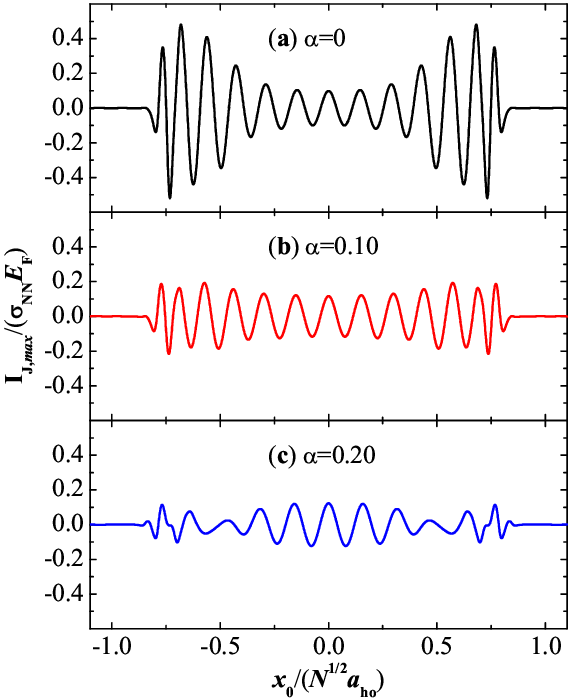}
\par\end{centering}

\caption{(Color online) Averaged maximum Josephson current at several shot to shot number 
fluctuations, $\alpha = 0$, $0.10$, and $0.20$. The BCS superfluid in the left well has a 
fixed atom number $N_L=N=128$, while the FFLO superfluid in the right well is subjected to a 
number fluctuation, with which the atom number $N_R$ is allowed to vary in the range $[N(1-\alpha), N(1+\alpha)]$. 
The two wells have the same interaction strength set by $\gamma _L = 1.6$. The spin polarization 
in the FFLO superfluid is fixed to $p=0.25$.}

\label{fig6} 
\end{figure}

We check in Fig. 6 how the FFLO signal is affected by the shot to shot fluctuations
in atom numbers at zero temperature. As a concrete example, we consider a BCS superfluid with a fixed number of atoms in the left well and 
a FFLO superfluid in the right well. The atom number in the FFLO superfluid is allowed to vary within 
a certain range. The curve in Fig. 6 shows the maximum Josephson current after averaging 
over 16 configurations. We find that our scheme is robust if the number fluctuation is 
less than $10 \%$, within the experimental resolution of measuring the atom number. In particular,
we find that the FFLO oscillation at the trap center persists up to the $30 \%$ fluctuation in 
atom numbers.

We finally discuss two issues concerning the temperature and interactions. (i) In 
contrast to the 3D case, the 1D FFLO state is notably stable in response to a nonzero 
temperature. In magnitude the critical temperature of the FFLO state is at the same 
order (\textit{i.e.}, a half or one third) of its unpolarized counterpart, 
$T_{BCS}\sim 4.54e^{-\pi ^2/2\gamma}T_F$, where $T_F$ is the Fermi temperature. 
Given an intermediate interaction $\gamma =1.6$, we estimate a critical temperature 
$T_{\text{FFLO}}\sim 0.10T_F$ at the trap center, which is well above the lowest temperature
reported so far. We anticipate a lower transition temperature inside the
weak link because of reduced density. However, it can be much enhanced
by increasing $\gamma$. (ii) In practice, $g_{1d}$ is parameterized by a 3D
scattering length $a_{3d}$, $g_{1d}=2\hbar ^2\omega _{\bot}a_{3d}/(1-Aa_{3d}/a_{\bot })$, 
where $a_{\bot }=\sqrt{\hbar /m\omega _{\bot}}$ and $A\simeq 1.0326$. The denominator 
indicates a confinement-induced Feshbach resonance that occurs when 
$a_{3d}\sim a_{\bot }$~\cite{cirth,ourcir}, as observed experimentally~\cite{cirexp}. 
For $^{6}$Li, using a Feshbach resonance the 3D scattering length $a_{3d}$ and hence the dimensionless
coupling constant $\gamma$ can be changed at will. 

\section{Conclusions}

In summary, we have proposed a scheme to realize the atomic Josephson junction to 
detect the possible existence of the exotic FFLO superfluids in quasi-one-dimensional 
atomic Fermi gases. Though there are several difficulties for realizing such a 
junction configuration, considering the rapid developments in cold-atom 
experiments, we anticipate that they will be overcome soon in the near future. Our 
proposal opens the possibility for creating ultracold Fermi gases for practical 
applications, such as precision measurement and interferometry.

\begin{acknowledgments}
We acknowledge fruitful discussions with Professor P. D. Drummond. 
This work was supported by the Australian Research Council (ARC) Centre 
of Excellence for Quantum-Atom Optics and the ARC Discovery Projects 
No. DP0984522 and No. DP0984637.
\end{acknowledgments}

\end{document}